\documentclass[doublecol]{epl2}
\usepackage{graphicx,amsmath,amssymb,amsfonts,latexsym,color,dcolumn,bm,epsfig,subfigure}
\usepackage[latin1]{inputenc}

\newcommand{\atwo}{a_{2\rm D}^{\rm eff}}
\newcommand{\rr}{{\bf r}}
\newcommand{\qq}{{\bf q}}
\newcommand{\RR}{{\bf R}}
\newcommand{\KK}{{\bf K}}
\newcommand{\ttt}{{\bf t}}
\newcommand{\HH}{\mathcal{H}}
\newcommand{\AAA}{\mathcal{A}}

\title{Matter Waves in Atomic Artificial Graphene}

\author{Nicola Bartolo \inst{1,2,3} \thanks{E-mail: \email{nicola.bartolo@univ-montp2.fr}}
\and Mauro Antezza\inst{1,2,4} \thanks{E-mail: \email{mauro.antezza@univ-montp2.fr}}}
\shortauthor{N. Bartolo \and M. Antezza}

\institute{                    
  \inst{1} Universit\'{e} Montpellier 2, Laboratoire Charles Coulomb UMR 5221 - F-34095 Montpellier, France\\
  \inst{2} CNRS, Laboratoire Charles Coulomb UMR 5221 - F-34095 Montpellier, France\\
  \inst{3} INO-CNR BEC Center and Dipartimento di Fisica, Universit\`{a} di Trento - I-38123 Povo, Italy\\
  \inst{4} Institut Universitaire de France - 103, bd Saint-Michel - F-75005 Paris, France
}
\pacs{03.75.-b}{Matter waves}
\pacs{81.05.ue}{Graphene}
\pacs{67.85.-d}{Ultracold gases, trapped gases}

\abstract{We present a new model to realize artificial {2D lattices with cold atoms investigating the atomic artificial} graphene: a 2D-confined matter wave is scattered by atoms of a second species trapped around the nodes of a honeycomb optical lattice. {The system allows an exact determination of the Green function, hence of the transport properties. The inter-species interaction can be tuned via the interplay between scattering length and confinements.} Band structure and density of states of {a periodic} lattice are derived for different values of the interaction strength. Emergence and features of Dirac cones are pointed out, together with the appearance of multiple gaps and a non-dispersive and isolated flat band. Robustness against finite-size and vacancies effects is numerically investigated.}

\begin{document}

\maketitle

\section{Introduction}

Due to the fundamental role of carbon in biological systems,
the investigations of physical properties of its allotropes
has always raised great interest.
This is the case for graphene, which is a flat monolayer of carbon atoms arranged
in a two-dimensional (2D) honeycomb lattice (HL) \cite{GeimNature07andCastroNetoRMP09}.
The experimental isolation of graphene in 2004 \cite{NovoselovScience04}
pushed out of the mere academic interest the study of this intriguing material,
with particular interest toward the outstanding transport properties of charge
carriers. These follow from the peculiar band structure of electrons:
conduction and valence bands touch in isolated point of $k$-space,
around which the energy-momentum dispersion relation is conical.
In such a scenario {Schr\"odinger's} equation fails to describe the particles' behavior,
ruled by a Dirac-like equation for massless fermions.
The group velocity $v_g\!\simeq\!10^6{\rm m/s}$ of these particles around
the Dirac cone plays the role of an effective speed-of-light of the charge carriers.
This intriguing scenario allows for the investigation of quantum electrodynamics
in benchtop experiments.

The interest towards these amazing properties led to the realization of several kinds of artificial graphene:
systems whose geometrical symmetries allow for the appearance of Dirac singularities {(for a recent review, see \cite{PoliniNature13})}.
{Among these we find nano-structured surfaces on which hexagonal patterns are impressed \cite{SinghaScience11,JacqminPRL14} and molecular graphene, obtained by accurately deposing molecules on a substrate \cite{GomesNature12}. Microwave analogs of graphene can also be implemented making use of HL of dielectric resonators \cite{BellecPRL13}.}
In this domain ultracold gases in optical lattices (OL) revealed themselves as a powerful and versatile tool due to the tunability of inter-particle interactions and lattice geometries \cite{ZhuPRL07andBlochNature12}.
Several theoretical approaches predict the existence of Dirac points in such kind of systems,
analyzing also their motion and merging acting on experimental parameters \cite{DietlPRL08, LimPRL108}.
These effects have been recently observed for a one-component ultracold Fermi gas \cite{TarruellNature12}.
The emergence of non-dispersive and non-isolated flat bands for cold-atoms {in honeycomb lattices} have also been predicted \cite{WuPRL07} and experimentally observed for polaritons \cite{JacqminPRL14}.

\begin{figure}[tb]
\includegraphics[width=0.45\textwidth]{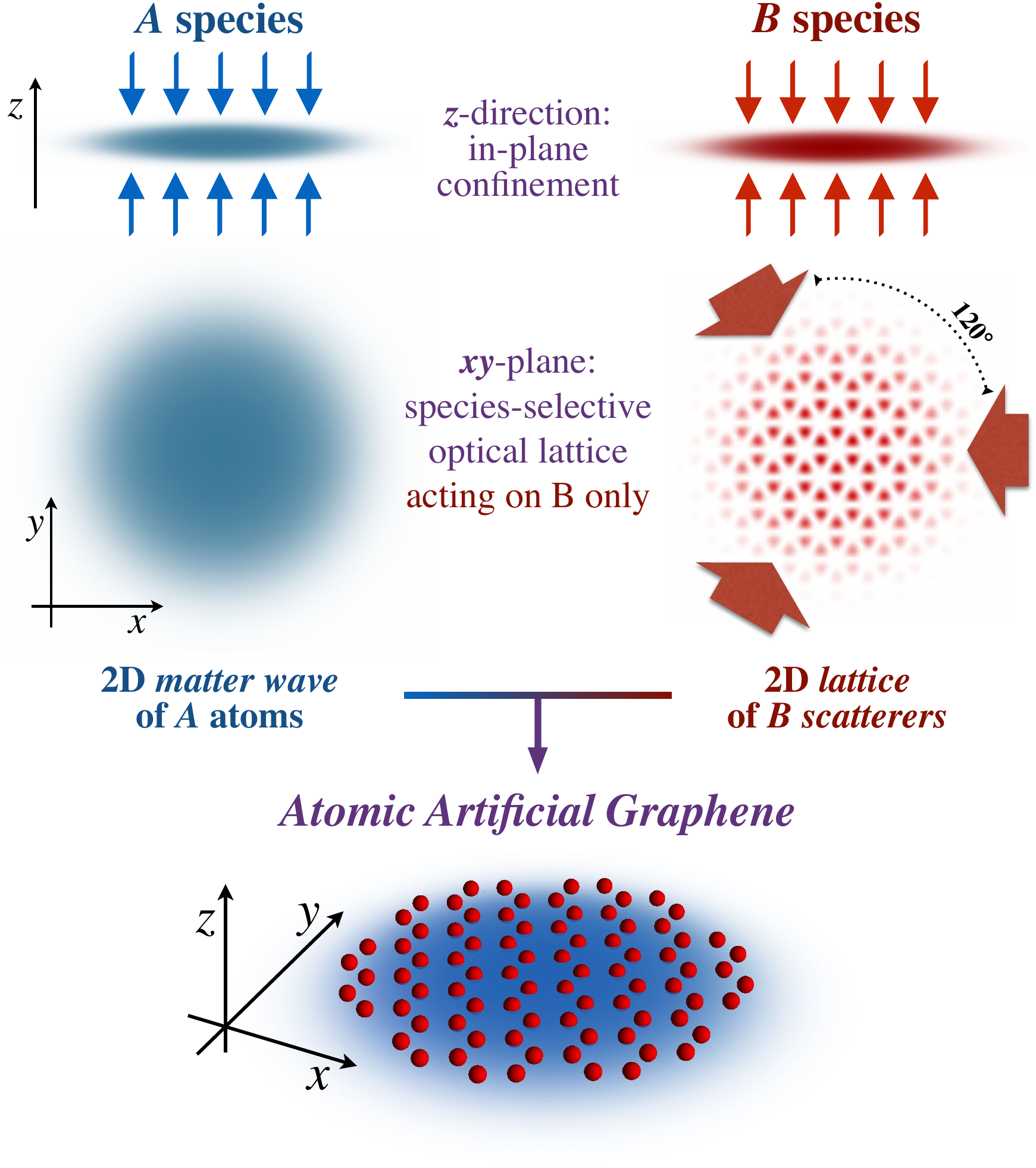}
\caption{(Color online)
Schematic representation of our model for the realization of atomic artificial graphene.
Two atomic species, namely $A$ (depicted in blue) and $B$ (depicted in red) are strongly confined on a plane.
Making use of a species-selective optical lattice in trine configuration $B$ atoms are trapped to form a 2D honeycomb lattice of point-like scatterers. In-plane confined $A$ atoms form a 2D matter wave which propagates through the artificial crystal.}
\label{FigSystem}\end{figure}

{In this letter we propose a general model for the realization of 2D artificial lattices with cold atoms.
We focus in particular on a novel kind of artificial graphene considering}
a two-species system in which a 2D-confined matter wave (MW) is subject to a periodic potential generated by point-like atomic scatterers pinned at the nodes of a honeycomb OL.
We refer to this system as Atomic Artificial Graphene (AAG) since, in contrast to previous proposal and realizations {involving cold atoms}, the periodic potential felt by the MW, made of $A$-species atoms (electrons in real graphene), is not a continuous optical potential but it is made by atoms of the species $B$ independently trapped in the vibrational ground state of deep parabolic lattice micro potentials (covalent atomic crystal in real graphene).
{A schematic representation of our system is proposed in fig.~\ref{FigSystem}.}
This model, recently suggested to realize disordered systems in 1D \cite{GavishPRL05}, 2D and 3D \cite{AntezzaPRA10}, due to point-like interactions allows an exact evaluation of the transport properties also for very large systems. Its lattice, whose  spacing is $\simeq\!0.5\mu$m ($>\!10^3$ times that of real graphene), can be easily tuned and deformed.
{Finally disorder, in the form of a fraction of random empty sites, can be naturally added.}

\section{{A general model}}
{Thanks to the experimental developments of the last decades, ultracold gases in optical lattices constitutes a perfect playground for the realization of quantum simulators of condensed matter systems \cite{ZhuPRL07andBlochNature12}. 
On one hand several techniques allow to tune the interaction between cold particles, for instance the $s$-wave  scattering length between atoms can be adjusted by an external magnetic field thanks to Feshbach resonances.
On the other hand optical lattices consent to create periodic or quasi-periodic potentials of arbitrary geometry by use of interfering laser beams. As an example three beams in trine configuration generate an honeycomb OL \cite{GrynbergPR01} (fig.~\ref{FigSystem}).
Recently a new holographic technique has been employed to realize arbitrary lattices of microtraps with exactly one atom per site \cite{NogrettearXiv14}, paving the way to the realization of \emph{tailored} OL in which geometry and lattice spacing can be tuned at will.
It is also possible to fix the frequency of a lattice in order to make it invisible to some atoms.
This species-selective OL have been first realized in \cite{LamporesiPRL10} on a mixture of $^{87}$Rb and $^{41}$K by adding a 1D OL whose frequency falls exactly in between two $^{87}$Rb resonances, so that the attractive and repulsive contribution of the optical potential cancel each other and only $^{41}$K results trapped in 2D sheets.
In such kind of system atoms of different species undergo the so-called mixed-dimension scattering, theoretically investigated for nD-3D mixtures \cite{MassignanPRA06andNishidaPRA10}. The process can be described in terms of an effective 3D scattering length which, depending on the trapping frequencies, experiences several confinement-induced resonances (observed in \cite{LamporesiPRL10}).

With current experimental techniques it is thus possible to implement our atomic artificial lattice in which $A$ and $B$ atoms are confined on a plane and a 2D selective OL is applied on $B$ atoms only (fig.~\ref{FigSystem}).
Properly tuning the lattice depth and the $B\!-\!B$ interaction, a Mott insulating phase with one atom per lattice site can be reached \cite{FisherPRB89, GreinerNature02} and controlled \cite{BakrScience10}.
We may also assume $A$ atoms to be non-interacting with themselves, a scenario realizable by using polarized fermions or bosons at zero scattering length.
Concerning the $A\!-\!B$ interaction, we have a 0D-2D scattering process for which, in analogy with \cite{MassignanPRA06andNishidaPRA10}, we can introduce an effective 2D scattering length $\atwo$ \cite{BartoloPrep}.}
We recall that for 2D systems, contrary to 3D and 1D cases, even at low energies the scattering amplitude stays $k$-dependent and only in the limit $\atwo\!\to\!0$ ($\atwo\!\to\!\infty$) we can consider the interaction to be overall weakly repulsive (attractive) \cite{OlshaniiPRL01andPricoupenkoJPB07}.
{The strong-interaction regime is reached for $k\atwo\!\sim\!1$
and a bound-state of energy $E_{\rm bs}\!=\!-(2/e^\gamma\atwo)^2$ exists ($\gamma\!\simeq\!0.577$ is the Euler-Mascheroni constant).}

We introduce our general theoretical approach by considering an $A$-atom 2D MW interacting with $N$ point-like
$B$-scatterers fixed at positions $\{\rr_i\}$, with $i\!=\!1,2,\cdots N$.
We can write the MW hamiltonian as that of a free $A$ atom, 
$\HH\!=\!-\frac{\hbar^2}{2m}\nabla_{\!\! 2\rm D}^2$, where $\nabla_{\!\! 2\rm D}^2$ is the
2D Laplace operator, and take into account the $A\!-\!B$ interaction by imposing the Bethe-Peierls 
contact conditions, \textit{i.e.} that the $A$ atom wave-function in the vicinity of the $i^{\rm th}$
scatterer behaves as
$\psi(\rr)\!\to\!
\frac{m}{\pi\hbar^2}\,   D_i\,   \ln\!\left(|\rr-\rr_i|/\atwo\right)
+O \left( |\rr-\rr_i| \right)$,
where $D_i$ is an arbitrary complex coefficient.
System's eigenstates correspond to poles of the Green's function that,
for point-like scatterers and MW source in $\rr_0$ is
\begin{equation}\label{EqGreen}
G(\rr,\rr_0)=g_0(\rr-\rr_0)+\sum_{i=1}^N D_i\, g_0(\rr-\rr_i).
\end{equation}
Here $g_0(\rr)\!=\!-(im/2\hbar^2)H_0^{(1)}(kr)$ is the
2D free space Green function that we would obtain in absence of scatterers
for $E\!=\!\hbar^2k^2/2m$, and $H_0^{(1)}$ is the Hankel function
of the $1^{\rm st}$ kind of index zero. For the wave-vector $k$ we
impose $k\!>\!0$ if $E\!>\!0$ and $k\!=\!i\kappa$ with $\kappa\!>0\!$ if $E\!<\!0$ (corresponding to bound states).
The $N$ coefficients $D_i$ defining $G$ in eq.~\eqref{EqGreen}
come from the solution of the {Schr\"odinger} equation, hence of the $N\!\times\!N$ complex linear
system  \cite{AntezzaPRA10}, 
\begin{equation}\label{EqLinearSystem}
\sum_{j=1}^N M_{ij} D_j =
-\frac{\pi\hbar^2}{m} g_0(\rr_i-\rr_0)
\qquad i=1,2,\cdots,N,
\end{equation}
with
\begin{equation}\label{EqDefM}
M_{ij}=
\begin{cases}
\frac{\pi\hbar^2}{m} g_0(\rr_i-\rr_j) & \rr_i\neq \rr_j \\
\ln\left(\frac{e^\gamma}{2}\,k\atwo\right)-i\frac{\pi}{2} & \rr_i=\rr_j.
\end{cases}
\end{equation}
It follows from eq.~\eqref{EqGreen} that $G$ would diverge if the
linear system~\eqref{EqLinearSystem} has no solution or, equivalently, if the
matrix $M$
admits zero as eigenvalue, \textit{i.e.} $\det(M)\!=\!0$.

{We can now consider the case of}
an arbitrary non-Bravais lattice made by two identical Bravais sub-lattices
{of primitive vectors ${\bf a}_1$ and ${\bf a}_2$ displaced by ${\bf t}$ with respect to each other.
Such a periodic structure results invariant under translation $\RR\!\in\!L$ with $L\!=\!\{n_1{\bf a}_1+n_2{\bf a}_2:n_1,n_2\in {\mathbb Z}\}$.
The reciprocal lattice vectors are defined as $\KK\!\in\!RL$ with $RL=\{n_1{\bf b}_1+n_2{\bf b}_2:n_1,n_2\in {\mathbb Z}\}$, where ${\bf b}_1$ and ${\bf b}_2$ satisfy ${\bf a}_i\cdot{\bf b}_j\!=\!2\pi\delta_{ij}$ ($i,j\!=\!1,2$).}
Bloch's theorem implies that only $D_i$ coefficients corresponding
to the same sub-lattice are correlated as $D_j\!=\!D_i\exp{[i\qq\cdot(\rr_i-\rr_j)]}$
\cite{AntezzaPRL09andAntezzaPRA09}, where $\qq$ is a wave-vector belonging to the first
Brillouin zone (FBZ).
Properly resorting to this property, and after some algebraic
manipulations of eqs.~\eqref{EqLinearSystem}-\eqref{EqDefM}, the condition for the existence
of an eigenstate results $\det(T)\!=\!0$,
where $T$ is a $2\!\times\!2$ matrix:
\begin{align}\label{EqChiReal}
&T_{11}=T_{22}=\ln\left(\frac{e^\gamma}{2}\,k\atwo\right)-i\frac{\pi}{2}
+\sum_{\RR\in L^*} \frac{\pi\hbar^2}{m} g_0(\RR) e^{i\qq\cdot\RR},
\nonumber\\
&T_{12}=T_{21}^*=\sum_{\RR\in L} \frac{\pi\hbar^2}{m} g_0(\RR+\ttt) e^{i\qq\cdot\RR},
\end{align}
where $L^*\!=\!L\setminus\{0\}$.
As usual, to obtain a faster convergence,
the sums in eqs.~\eqref{EqChiReal} can be rewritten
in the reciprocal space using Poisson's identity: 
\begin{align}\label{EqChiReciprocal}
&T_{11}=T_{22}=
\ln\left(\frac{e^\gamma}{2}\right)+\alpha+ C_\infty + \frac{2\pi}{\AAA}\frac{1}{k^2-q^2}
\nonumber\\
&\quad
+\frac{2\pi}{\AAA} \sum_{\KK\in RL^*} \left(\frac{1}{k^2-|\KK-\qq|^2}
+\frac{1}{K^2} \right) ,
\nonumber\\
&T_{12}=T_{21}^*=\frac{2\pi}{\AAA} \sum_{\KK\in RL} \frac{e^{i(\KK-\qq)\cdot\ttt}}{k^2-|\KK-\qq|^2},
\end{align}
where $\AAA$ is the area of the real-space unit cell.
Here the diagonal terms are explicitly real and the solution of $T_{11}\!=\!0$ provides the spectrum for one of the Bravais sub-lattices. $C_\infty$ is a coefficient depending only on the geometry of the Bravais sub-lattice \cite{NoteCinfty}.
For a triangular sub-lattice, building-block of graphene, $C_\infty\!\simeq\!0.959662$.
The interaction strength $\alpha\!=\!\log(\atwo/a)$ ($a$ being the lattice spacing) enters only through diagonal terms, and the band structure can be directly calculated by solving in $\qq$ and $E$ the equations $T_{11}(\alpha\!=\!0)\pm|T_{12}|\!=\!-\alpha$. It is worth noticing that this equation, together with eqs.~\eqref{EqChiReal}-\eqref{EqChiReciprocal}, is one of the main results of the paper and it is valid for arbitrary non-Bravais lattices with two atoms per unit cell. 

\begin{figure}[htb]
\includegraphics[width=0.48\textwidth]{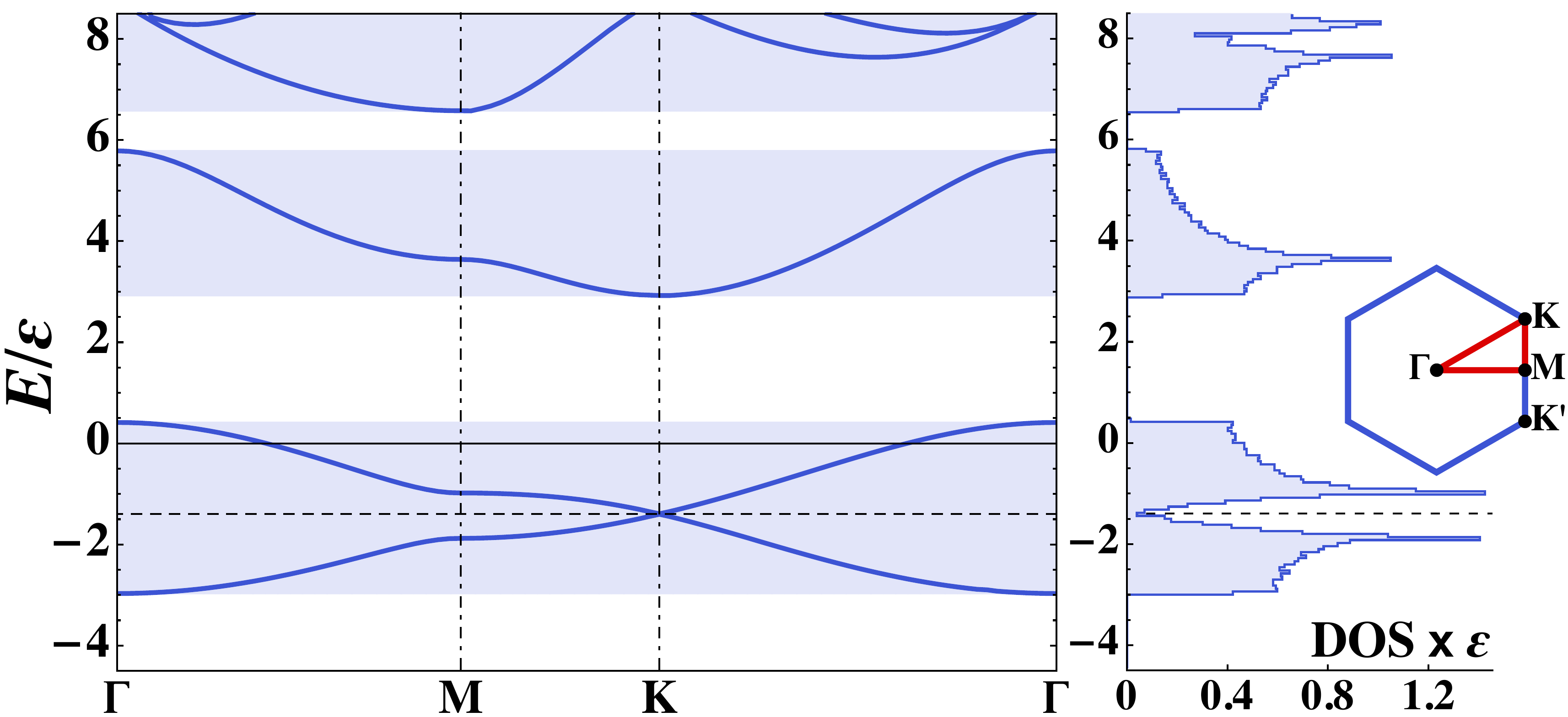}
\caption{(Color online)
Band structure (left) and DOS (right) for the AAG 
at $\alpha\!=\!-0.5$. The five lowest energy bands are
evaluated along the irreducible symmetry path in the inset.
To obtain the DOS, the energies of the same bands have been evaluated
in $N_{\!s}\!\simeq\!3300$ points sampled inside the red path.
For each bin of the histogram (of width $\delta E\!=\!0.06\varepsilon$, $\varepsilon\!=\!\hbar^2/ma^2$)
the band normalized density of states is ${\rm DOS}=N_{\!E}/N_{\!s}\,\delta E$, where $N_{\!E}$ is the
number of sampled energies falling in the bin.}
\label{FigSpecter}\end{figure}

\section{{Dirac cones and flat band in AAG}}
{We specify now to the exemplary case of the AAG.
$B$-scatterers form thus an HL in which $a$ is the distance between an atom and its three nearest neighbors.
This arrangement corresponds to a non-Bravais triangular lattice of primitive vectors ${\bf a}_1\!=\!\frac{3 a}{2}(1,1/\sqrt{3})$ and ${\bf a}_2\!=\!\frac{3 a}{2}(1,-1/\sqrt{3})$ with relative displacement $\ttt\!=\!a\,(1,0)$.
Each unit cell have area $\AAA\!=\!3\sqrt{3}a^2/2$
and the reciprocal primitive vectors result ${\bf b}_1\!=\!\frac{2\pi}{3 a}(1,\sqrt{3})$
and ${\bf b}_2\!=\!\frac{2\pi}{3 a}(1,-\sqrt{3})$.}

In fig.~\ref{FigSpecter} we show a typical band structure of the AAG evaluated along an high-symmetry path in the $k$-space of the HL,  for $\alpha\!=\!-0.5$.
For sake of completeness it is also possible to numerically
evaluate the density of states (DOS) of the system by sampling the eigenvalues $E$ over the FBZ.
The results are shown in fig.~\ref{FigSpecter} beside the corresponding band structure.
We find a multi-gapped spectrum in which  the two lowest bands touch at $E\!=\!E_D\!\simeq\!-1.40\varepsilon$ ($\varepsilon\!=\!\hbar^2/ma^2$) in the inequivalent points ${\rm K, K}'\!=\!2\pi/3a(1,\pm1/\sqrt{3})$ of the FBZ, giving rise to the  typical Dirac-cone dispersion relation.
Correspondingly the DOS goes to zero.
At higher energies we find an isolated band followed by a continuum of states.

\begin{figure}[htb]
\includegraphics[width=0.48\textwidth]{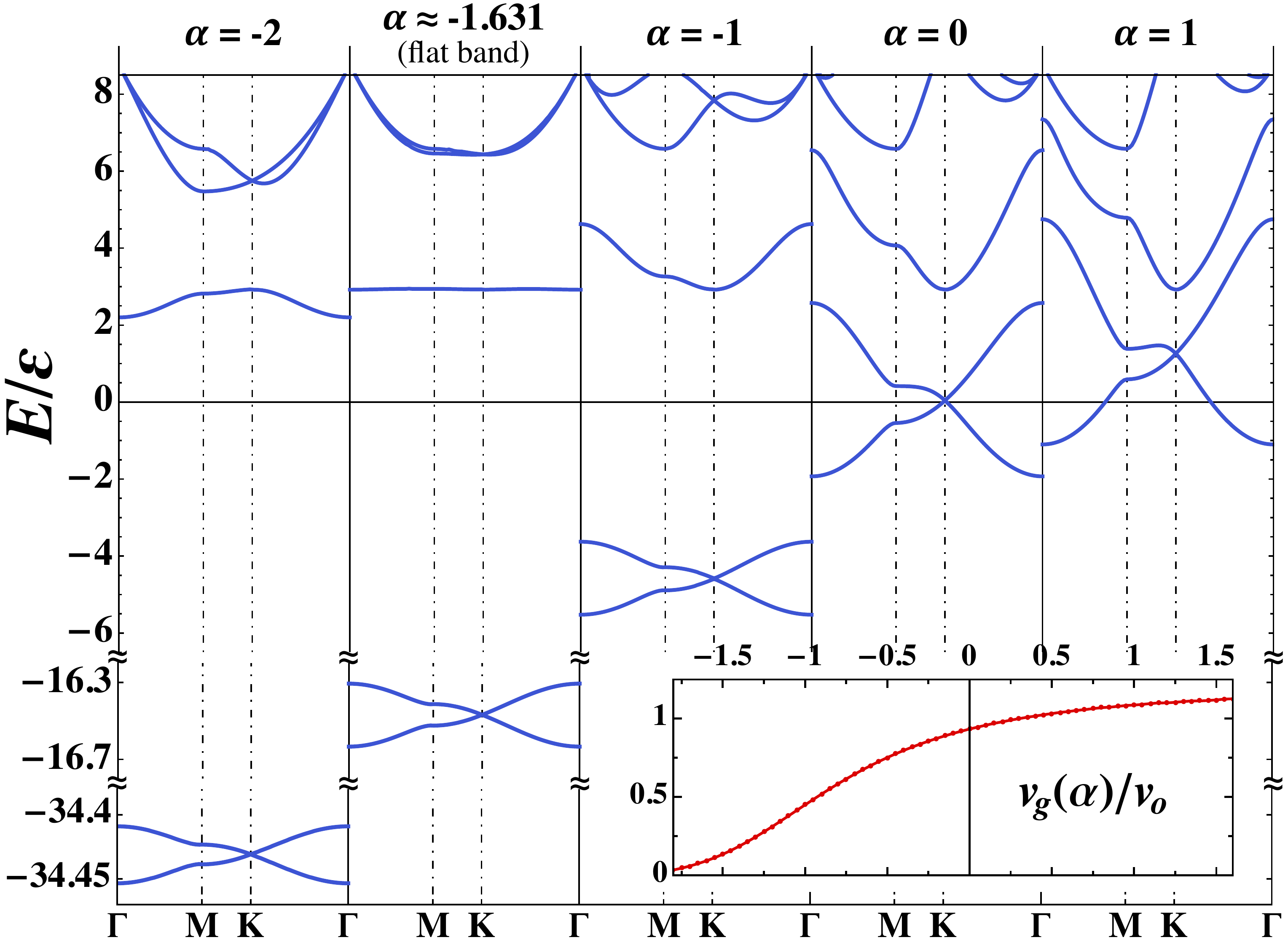}
\caption{(Color online)
Band structure of an AAG for different values of
the interaction-parameter $\alpha$. For $\alpha\simeq1.63$ an isolated flat band appears.
Here $\varepsilon\!=\!\hbar^2/ma^2$.
Inset: Modulus of the group velocity $v_g$ for an $A$-atoms MW
around the Dirac's cone as a function of $\alpha$.
Velocities are normalized on $v_o\!=\!\hbar/ma$
($\simeq\!1.5{\rm mm/s}$ for MW of $^{87}$Rb
atoms in an OL with $a\!=\!500{\rm nm}$).}
\label{FigSpecterCompared}\end{figure}

The remarkable tunability of the AAG emerges from fig.~\ref{FigSpecterCompared}, in which band structures corresponding to different values of $\alpha$ are compared.
We see how the gaps can be modulated and closed tuning the interaction strength.
The two lowest bands support the Dirac cone,  which for $\alpha\!\lesssim\!0$ moves to negative energies, making the relativistic physics played now by states bounded in the system\cite{NoteDiracBound}.
The same two bands go fast deeper and flatter by reducing $\alpha$, but they still touch only at K and K$'$. It follows that the slope of the Dirac-cones' walls decreases, leading to a reduction of the group velocity $v_g$ around $E_D$ (see inset of fig.~\ref{FigSpecterCompared}) remarkably down to $v_g\!\lesssim\!1{\rm mm/s}$, \textit{i.e.} {$10^{-9}$} the value for real graphene.

It is worth stressing that for $-2\!<\!\alpha\!<\!-1$ the third band changes concavity, becoming completely flat around $E\!\simeq\!2.92\varepsilon$ for $\alpha\!\simeq\!-1.63$.
Remarkably, differently from previously investigated {graphene-like} systems \cite{WuPRL07, JacqminPRL14}, here the flat band results also isolated.
On this non-dispersive band $A$ atoms behave as particles of infinite mass and any state of the matter-wave is localized.
The flat band would drastically enhance any effect of the $A\!-\!A$ interaction (that here we assumed to be zero) leading to the emergence of strongly correlated phases {and frustration} \cite{WuPRB08}.

Beside allowing to tune interactions, the use of cold atoms in optical lattices offers a large experimental control on the potential landscape. For one-component artificial graphene a manipulation of the optical potential can lead to a displacement of the Dirac cones within the FBZ, eventually resulting in their merging and disappearance \cite{DietlPRL08,LimPRL108}.
In the case of our two-component AAG, a distorsion in the honeycomb arrangement of $B$ scatterers results in similar effects.

\begin{figure}[htb]
\includegraphics[width=0.48\textwidth]{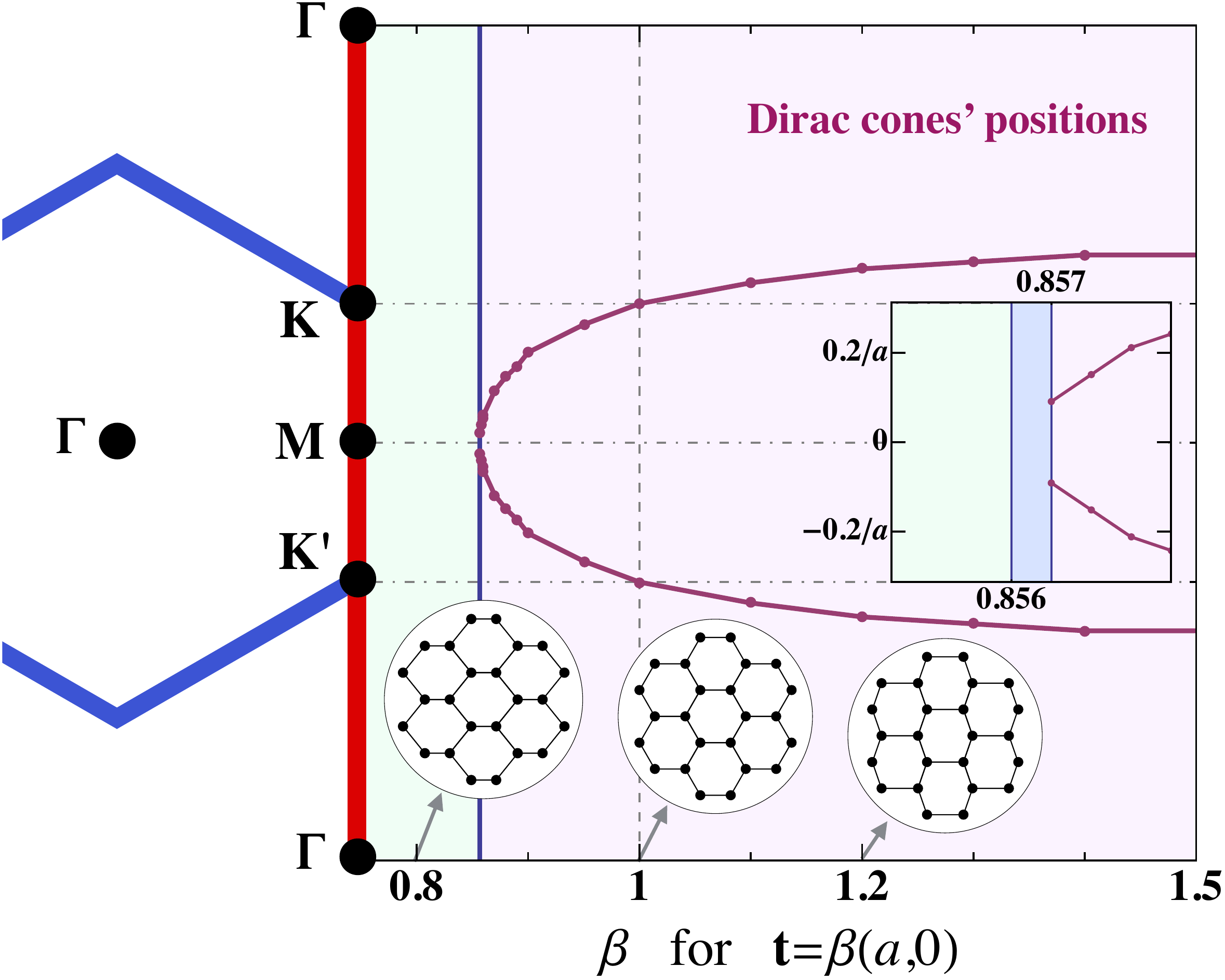}
\caption{(Color online)
For an AAG with displacement vector $\ttt\!=\!\beta(a,0)$, we show as a function of the parameter $\beta$ the position of the Dirac cones, if existing, along the $\Gamma\!-\!{\rm K'}\!-\!{\rm M}\!-\!{\rm K}\!-\!\Gamma$ path (highlighted in red to the left on a schematic representation of the FBZ).
Present results have been obtained for $\alpha\!=\!-0.6$.
Right-side inset: Zoom-in around the merging region. Results show that Dirac points meet and cancel each other in $\rm M$ for $\beta\!\sim\!0.86$.
Bottom-part circular insets: Real-space arrangement of the scatterers in the AAG for the indicated values of $\beta$: 0.8, 1 (\textit{i.e.} undistorted) and 1.2 from left to right.}
\label{FigDiracConesMotion}
\end{figure}

We investigated the motion and merging of Dirac points by considering, fixed $\alpha\!=\!-0.6$, different displacement vectors of the form $\ttt\!=\!\beta(a,0)$. Results are shown in fig.~\ref{FigDiracConesMotion}, where the positions of the cones along the $\Gamma\!-\!{\rm K'}\!-\!{\rm M}\!-\!{\rm K}\!-\!\Gamma$ path are plotted taking $\beta$ as parameter.
For $\beta\!=\!1$ we are in the case of undistorted graphene, and we find again that the Dirac points lay at $\rm K$ and $\rm K'$. The cones depart vertically by increasing $\beta$, \textit{i.e.} by pushing horizontally apart the scatterers within the unit cell.
On the other hand for $\beta\!<\!1$ the cones approach each other and merge in $\rm M$ for $0.856\!<\!\beta\!<\!0.857$ (see zoom-in of fig.~\ref{FigDiracConesMotion}). The dispersion relation at the merging point shows {the typical semi-Dirac} behavior: it is parabolic along the merging direction but stays linear along the perpendicular one \cite{LimPRL108}.
For smaller values of $\beta$ a gap is opened and Dirac points finally disappear.
These features remains qualitatively the same if $\alpha$ is set to a different value.
As a technical comment we point out that for exactly $\beta\!=\!1.5$ the lattice degenerates in a Bravais rectangular one and the corresponding equation should be solved.

\section{Effects of finite-size and vacancies}
Up to now we dealt with an ideal infinite system. 
For both theoretical and practical needs related to experiments, it is crucial to understand how and when the 
features of an infinite perfect AAG are modified if both finite-size and vacancy effects are considered.
It is typically possible to manipulate atomic clouds in OLs
extending over $\sim\!60$ sites per direction \cite{GreinerNature02},
for a total of $\sim\!10^3$ available traps for the scatterers in a 2D system.
Given a set of $N$ positions for the $B$-scatterers,
an eigenstate of the system exists for each couple $\{\alpha,E\}$
such that $\det(M)\!=\!0$, being $M$ the $N\!\times\!N$ matrix
defined in eq.~\eqref{EqDefM}.
The interaction-dependent term results isolated by writing $M\!=\!M_0+{\rm}I\alpha$,
{where $M_0\!=\!M(\alpha\!=\!0)$ does not depend on $\alpha$ and $I$ is the 2$\times$2 identity matrix}.
This reduces the condition for the existence of an eigenstate to $m_{0i}\!=\!-\alpha$,
where $m_{0i}$ is the $i^{\rm th}$ eigenvalue of $M_0(E)$.
At negative energies the eigenstates are bound states for which the wave is trapped inside the gas of scatters  {in virtue of the $A\!-\!B$ interaction only}.
Correspondingly the matrix $M_0$ and its eigenvalues are real and the conditions $m_{0i}(E_{\rm bs})\!=\!-\alpha$ give exactly the real bound states energies $E_{\rm bs}$.
At positive energies the problem is more delicate since the poles $z$ of the Green function~\eqref{EqGreen}, analytically continued to the lower half complex plane, are  solutions of $m_{0i}(z)\!=\!-\alpha$, with $z\!=\!E-i\hbar/2\tau$ where $E$ and $\tau$ are the energy of the state and its lifetime inside the scattering region of radius $R$. For a large enough atomic lattice, and for quasi-Bloch bulk states, lifetime scales as $\tau\!\propto\!R/v_{g}$.

\begin{figure}[htb]
\includegraphics[width=.48\textwidth]{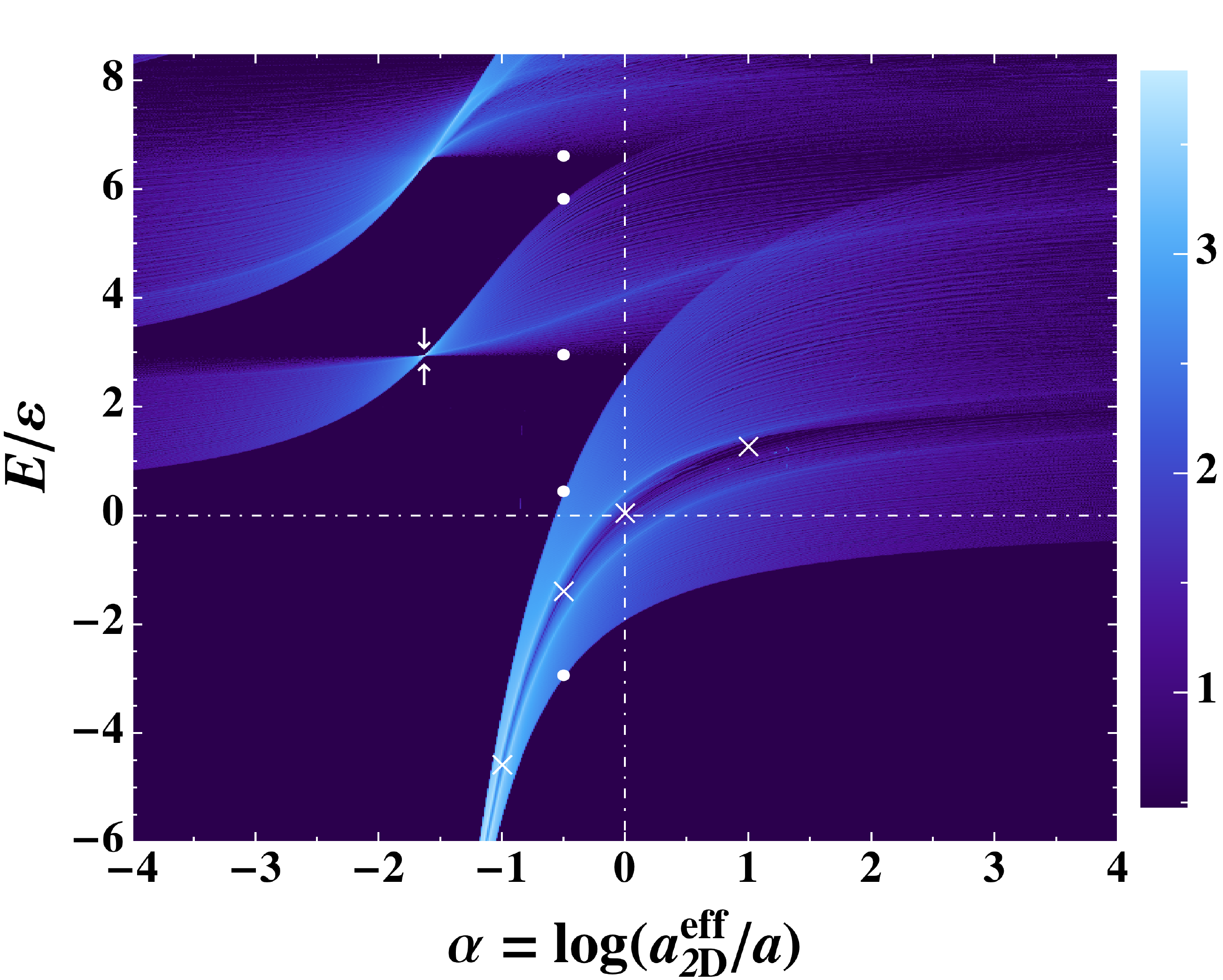}
\caption{(Color online) DOS per scatterer in the plane
$[E/\varepsilon,\alpha]$ for a system of $N\!=\!2167$ scatterers
arranged in a HL inside a disk of radius $R\!=\!30a$.
The energies are discretized with a step of $0.0025\varepsilon$, $\varepsilon\!=\!\hbar^2/ma^2$.
For given $E\!<\!0$ all the $N$ solutions of $m_{0i}(E)\!=\!-\alpha$ are selected.
For $E\!>\!0$ the sampled values are used as Newton's method starting points to find
the $N$ values of $z$ solving $m_{0i}(z)\!=\!-\alpha$. Here we select only quasi-Bloch bulk states, verifying $\tau\!\gtrsim\!5\hbar/\varepsilon\simeq R/{\rm max}(v_{g})$ (\textit{i.e.} $\tau\!\gtrsim\!2$ms for $^{87}$Rb
atoms MW with $a\!=\!500{\rm nm}$). 
The color-map is applied to 
$\log_{10}(\frac{N_{\rm eig}}{N}\frac{\varepsilon}{\delta\alpha\,\delta E})$,
where $N_{\rm eig}$ is the number of selected eigenstates within a
rectangular bin of area $\delta\alpha\,\delta E$
($\delta\alpha\!=\!0.0125$ and $\delta E\!=\!0.0125\varepsilon$).
$\times$, $\bullet$, and arrows indicate, respectively, the positions
of Dirac cones, gap boundaries, and isolated flat band as expected from the
analysis of an infinite system (figs.~\ref{FigSpecter} and \ref{FigSpecterCompared}).}
\label{FigFiniteSize}\end{figure}

In fig.~\ref{FigFiniteSize} we present the DOS as a function
of $E$ and $\alpha$ for a set of $\sim\!2000$ scatterers.
The features of the infinite system are already well reproduced,
as can be inferred from a comparison with figs.~\ref{FigSpecter} and
\ref{FigSpecterCompared}.
The large {dark} areas, in which no states are allowed, exactly correspond to gaps
in the infinite system. The expected boundaries of the gaps for $\alpha\!=\!-0.5$
are marked by $\bullet$ in fig.~\ref{FigFiniteSize}.
The fingerprint of the Dirac cone can be recognized in the thin dark
region separating the two lowest bands. The expected positions of the
cones, as deduced from the infinite-system results, are marked by $\times$. 
The existence of a flat band is confirmed in the finite-size system,
and its position is in perfect agreement with predictions from the
ideal HL (see arrows in fig.~\ref{FigFiniteSize}, top-left).
As a final remark we point out that in the weak-interaction limit,
\textit{i.e.} for $|\alpha|\!\gg\!1$, no states are allowed for $E\!<\!0$
while the DOS tends to a constant for $E\!>\!0$,
as expected for a free MW in 2D.

\begin{figure}[htb]
\includegraphics[width=.48\textwidth]{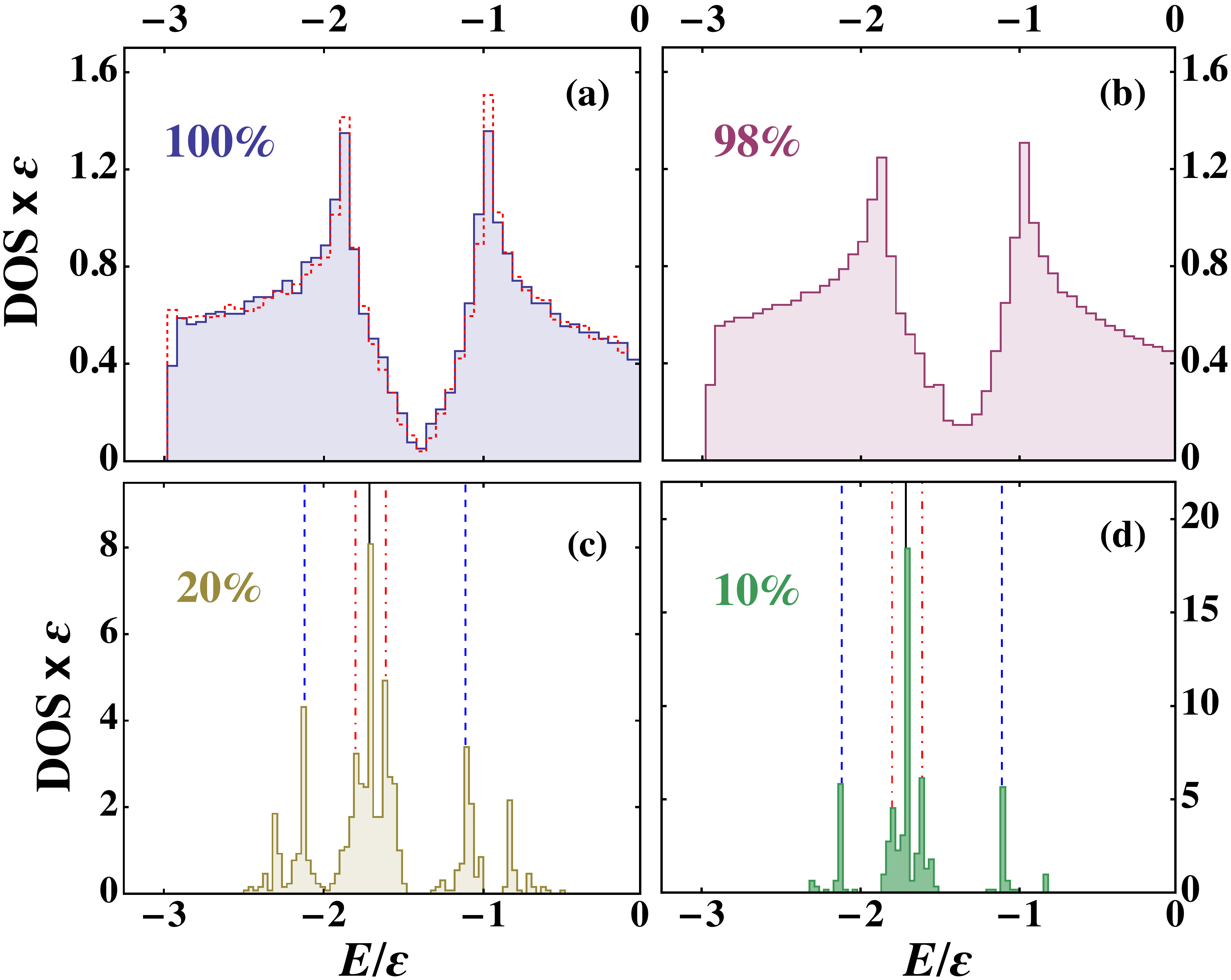}
\caption{(Color online)
DOS per scatterer for an HL with $N\!=\!3870$ sites, inside a circle of radius $R\!=\!40a$,
randomly occupied with filling factor 100\% (a), 98\% (b), 20\% (c), 10\% (d).
The eigenstates are evaluated as for fig.~\ref{FigFiniteSize} fixing $\alpha\!=\!-0.5$.
Histograms are obtained for a bin size $\delta E\!=\!0.06\varepsilon$ in (a)-(b)
and $\delta E\!=\!0.03\varepsilon$ in (c)-(d), with $\varepsilon\!=\!\hbar^2/ma^2$.
In order to facilitate the comparison,
the histograms are normalized as the corresponding
 DOS of the infinite periodic system
(dashed red histogram in (a)), \textit{i.e.} to $\sim\!1.82$.
The vertical lines in (c)-(d) show the energies of few-body bound states:
$AB$ dimer (solid black), $AB_2$ trimer with $B$ atoms separated
by $a$ (dashed blue) and $a\sqrt{3}$ (dot-dashed red).
}\label{FigDisorder}\end{figure}

{In the field of graphene simulation a large interest is devoted to the realization of disorder and the analysis of its effects \cite{PoliniNature13,BellecPRL13}. Our AAG naturally offers the possibility to introduce it in the form of a fraction of empty sites \cite{AntezzaPRA13}. By just loading the species-selective OL below the unitary filling one would end up with random vacancies in the HL.}
In fig.~\ref{FigDisorder} we present the DOS at negative energies
for different filling factors of an OL of $\sim\!4000$ sites, fixed $\alpha\!=\!-0.5$.
For a fully occupied HL (filing 100\%), in panel (a) the finite-system DOS is compared to the corresponding quantity for the infinite one, showing {again} the robustness of the Dirac cones with respect to the finite size of the system. 
Panel (b) shows the combined effect of a lattice of finite size and of $2\%$ unoccupied lattice sites.
The behavior is qualitatively the same as in panel (a), while the central minimum is higher but still clearly visible.
It starts disappearing when the fraction of empty sites is further increased. 
When the filling factor drastically decreases to $20\%$-$10\%$ (panels (c)-(d)), the systems becomes more and more disordered \cite{AntezzaPRA10}, and few-body effects start playing a crucial role, giving rise to strong peaks in the
DOS around the energies of $AB$ dimers and $AB_{2}$ trimers.
{The existence of disorder-localized states in such kind of 2D systems has been already investigated in \cite{AntezzaPRA10}.}
It is worth notice that the possibility of having an imperfect filling due to a fraction of empty sites is not available in one-species graphene realizations \cite{TarruellNature12}, and makes this system considerably richer toward the realization of quantum simulators of disordered graphene {with cold atoms}.

\section{Conclusions}
We presented a new model for the realization of {bi-dimensional artificial lattices with cold atoms}, investigating the properties of artificial atomic graphene.
For an ideal infinite system the band structure and DOS point out the existence of remarkable features: tunable multiple gaps, Dirac cones for bound states, reduced group velocity, and completely flat and isolated bands.
We also showed that Dirac points move and merge if the honeycomb structure is distorted.
The robustness of these features has been proved for finite systems of experimentally reasonable size, including the presence of vacancies in the lattice.
More disordered configurations, corresponding to large number of vacancies, show the emergence of
few-body physics. This system results then not only promising to mimic graphene physics on a benchtop scale,
but also to quantum simulate and predict the physics of 2D strongly correlated systems {in arbitrary geometries}.
In this direction, an interesting development would be the introduction of an effective $p$-wave $A\!-\!B$
interaction, as well as the $A\!-\!A$ one.
Furthermore the general approach presented is suitable for the analysis of other 2D geometries, with particular attention to disordered systems, not easily implementable with present one-species {cold-atoms} models.


\begin{thebibliography}{99}

\bibitem{GeimNature07andCastroNetoRMP09}
  \Name{Geim A. K.\and Novoselov K. S.}
  \REVIEW{Nature materials}{6}{2007}{183};
  \Name{Castro Neto A. H., Guinea F., Peres N. M. R., Novoselov K. S. \and Geim A. K.}
  \REVIEW{Rev. Mod. Phys.}{81}{2009}{109}.

\bibitem{NovoselovScience04}
  \Name{Novoselov K. S., Geim A. K., Morozov S. V., Jiang D., Zhang Y., Dubonos S. V., Gregorieva I. V. \and Firsov A. A.}
  \REVIEW{Science}{306}{2004}{666}.

\bibitem{PoliniNature13}
  \Name{Polini M., Guinea F., Lewenstein M., Manoharan H. C. \and Pellegrini V.}
  \REVIEW{Nature nanotechnology}{8}{2013}{625}.
  
{\bibitem{SinghaScience11}
  \Name{Singha A., Gibertini M., Karmakar B., Yuan S., Polini M., Vignale G., Katsnelson M. I., Pinczuk A., Pfeifer L. N., West K. W. \and Pellegrini V.}
  \REVIEW{Science}{332}{2011}{1176}.}
  
\bibitem{JacqminPRL14}
  \Name{Jacqmin T., Carusotto I., Sagnes I., Abbarchi M., Solnyshkov D. D., Malpuech G., Galopin E., Lema\^{i}tre A., Bloch J. \and Amo A.}
  \REVIEW{Phys. Rev. Lett.}{112}{2014}{116402}.

{\bibitem{GomesNature12}
  \Name{Gomes K. K., Mar W., Ko W., Guinea F. \and Manoharan H. C.}
  \REVIEW{Nature}{483}{2012}{306}.}
  
{\bibitem{BellecPRL13}
  \Name{Bellec M., Kuhl U., Montambaux G. \and Mortessagne F.}
  \REVIEW{Phys. Rev. Lett.}{110}{2013}{033902};
  \Name{Barkhofen S., Bellec M., Kuhl U. \and Mortessagne F.}
  \REVIEW{Phys. Rev. B}{87}{2013}{035101}.}

\bibitem{ZhuPRL07andBlochNature12}
  \Name{Zhu S.-L., Wang B. \and Duan L.-M.}
  \REVIEW{Phys. Rev. Lett.}{98}{2007}{260402};
  \Name{Bloch I., Dalibard J. \and Nascimb\`ene S.}
  \REVIEW{Nature Physics}{8}{2012}{267}.

{\bibitem{DietlPRL08}
  \Name{Dietl P., Pi\'echon F. \and Montambaux G.}
  \REVIEW{Phys. Rev. Lett.}{100}{2008}{236405}.}

{\bibitem{LimPRL108}
  \Name{Lim L.-K., Fuchs J.-N. \and Montanbaux G.}
  \REVIEW{Phys. Rev. Lett.}{108}{2012}{175303}.}

\bibitem{TarruellNature12}
  \Name{Tarruell L., Greif D., Uehlinger T., Jotzu G. \and Esslinger T.}
  \REVIEW{Nature}{483}{2012}{302}.

\bibitem{WuPRL07}
  \Name{Wu C., Bergman D., Balents L. \and Das Sarma S.}
  \REVIEW{Phys. Rev. Lett.}{99}{2007}{070401}.

\bibitem{GavishPRL05}
  \Name{Gavish U. \and Castin Y.}
  \REVIEW{Phys. Rev. Lett.}{95}{2005}{020401}.

\bibitem{AntezzaPRA10}
  \Name{Antezza M., Castin Y. \and Hutchinson D. A. W.}
  \REVIEW{Phys. Rev. A}{82}{2010}{043602}.

{\bibitem{GrynbergPR01}
  \Name{Grynberg G. \and Robilliard C.}
  \REVIEW{Phys. Rep.}{355}{2001}{335}.}
  
\bibitem{NogrettearXiv14}
  \Name{Nogrette F., Labuhn H., Ravets S., Barredo D., B\'{e}guin L., Vernier A., Lahaye T. \and Browaeyes A.}
  \REVIEW{Phys. Rev. X}{4}{2014}{021034}.

\bibitem{LamporesiPRL10}
  \Name{Lamporesi G., Catani J., Barontini G., Nishida Y., Inguscio M. \and Minardi F.}
  \REVIEW{Phys. Rev. Lett.}{104}{2010}{153202}.

\bibitem{MassignanPRA06andNishidaPRA10}
  \Name{Massignan P. \and Castin Y.}
  \REVIEW{Phys. Rev. A}{74}{2006}{013616};
  \Name{Nishida Y. \and Tan S.}
  \REVIEW{Phys. Rev. A}{82}{2010}{062713}.

\bibitem{FisherPRB89}
  \Name{Fisher M. P. A., Weichman P. B., Grinstein G. \and Fisher D. S.}
  \REVIEW{Phys. Rev. B}{40}{1989}{546}.

\bibitem{GreinerNature02}
  \Name{Greiner M., Mandel O., Esslinger T., H\"{a}nsch T. W. \and Bloch I.}
  \REVIEW{Nature}{415}{2002}{40}.

\bibitem{BakrScience10}
  \Name{Bakr W. S., Peng A., Tai M. E., Ma R., Simon J., Gillen J. I., F\"{o}lling S., Pollet L. \and Greiner M.}
  \REVIEW{Science}{329}{2010}{547}.

{
\bibitem{BartoloPrep}
  A detailed study of the dependence of $\atwo$ on the free scattering length, atomic masses, and trapping frequencies will be presented elsewhere.
For the purposes of this letter we simply assume that $\atwo$ can be experimentally tuned at will in its range of existence $[0,\infty[$ preserving the validity of the point-like approximation for $B$ scatterers.}

\bibitem{OlshaniiPRL01andPricoupenkoJPB07}
  \Name{Olshanii M. \and Pricoupenko L.}
  \REVIEW{Phys. Rev. Lett.}{88}{2001}{010402};
  \Name{Pricoupenko L. \and Olshanii M.}
  \REVIEW{J. Phys. B}{40}{2007}{2065}.

\bibitem{AntezzaPRL09andAntezzaPRA09}
  \Name{Antezza M. \and Castin Y.}
  \REVIEW{Phys. Rev. Lett.}{103}{2009}{123903};
  \Name{Antezza M. \and Castin Y.}
  \REVIEW{Phys. Rev. A}{80}{2009}{013816}.

\bibitem{NoteCinfty}
{Poisson's formula implies the evaluation of a principal value integral which can be overcome introducing an auxiliary ultra-violet cut-off. For a circular cut-off $k_{\rm uv}$ this leads to the definition of 
$C_{\rm uv}\!=\!\log(k_{\rm uv} a)-\frac{2\pi}{A}\sum^{'}\frac{1}{K^2}$,
where the sum runs over $\{\KK\in RL^*: {K < k_{\rm uv}}\}$, and whose
limit for $k_{\rm uv}\!\to\!\infty$ defines the quantity $C_\infty$.}
  
\bibitem{NoteDiracBound}
{The presence of 2D bound-states for the $A$ atoms localized around the $B$ scatterers makes possible the existence of Dirac cones also for $\alpha\!\ll\!-1$.}  
  
\bibitem{WuPRB08}
  \Name{Wu C. \and Das Sarma S.}
  \REVIEW{Phys. Rev. B}{77}{2008}{235107}.
  
\bibitem{AntezzaPRA13}
  \Name{Antezza M. \and Castin Y.}
  \REVIEW{Phys. Rev. A}{88}{2013}{033844}.

\end{thebibliography}
\end{document}